\documentclass[sigconf,nonacm]{acmart}

\usepackage{multirow}
\usepackage{algorithm}
\usepackage{algpseudocode}
\usepackage{textcomp}
\graphicspath{{figures/}}
\algnewcommand\algorithmicparfor{\textbf{parallel for}}
\algnewcommand\algorithmicendparfor{\textbf{end for}}
\algdef{SE}[PARFOR]{ParFor}{EndParFor}[1]{\algorithmicparfor\ #1\ \textbf{do}}{\algorithmicendparfor}

\newcommand{\ktree}{\texorpdfstring{$k^2$-tree}{k2-tree}}

\newcommand{\bigoh}{\mathcal{O}}
\newcommand{\popc}{\mathrm{popcount}}
\newcommand{\GB}{GB10}
\newcommand{\bin}[1]{(#1)_2}
\newcommand{\TO}{\ensuremath{\mathcal{T}}}
\newcommand{\CF}{\ensuremath{\mathcal{C}}}

\usepackage{cleveref}

\begin{document}

\title{A GPU Engine for Sparse Matrix-Vector Products on \texorpdfstring{$k^2$}{k2}-Tree-Compressed Graphs}

\author{Francesco Tosoni}
\affiliation{%
  \department{L'EMbeDS}
  \institution{Sant'Anna School of Advanced Studies}
  \city{Pisa}
  \country{Italy}}
\additionalaffiliation{%
  \institution{University of Pisa}
  \department{Department of Computer Science}
  \city{Pisa}
  \country{Italy}}
\email{Francesco.Tosoni@santannapisa.it}
\orcid{0000-0001-8457-3866}

\author{Gabriele Mencagli}
\affiliation{%
  \department{Department of Computer Science}
  \institution{University of Pisa}
  \city{Pisa}
  \country{Italy}}
\email{gabriele.mencagli@unipi.it}
\orcid{0000-0002-6263-7723}

\renewcommand{\shortauthors}{Tosoni and Mencagli}

\begin{abstract}
The \ktree{} is a compact representation of Boolean matrices: it stores no values and no coordinates, and serves rows and columns alike. On a GPU, however, finding the children of a node takes $\mathrm{rank}$ queries that depend on one another, so a node-by-node descent spends its time waiting on memory. We present a GPU engine for sparse matrix--vector multiplication (SpMV) on \ktree{}s. Since the positions of the children are the same for every vector, the engine visits the tree one level at a time and takes them from a prefix sum or from a directory sampled every 256 nodes, and working in tiles fixes the device memory of a product before it starts. One traversal yields products on either side, for one vector or a batch, over the plus-times, Boolean and tropical semirings. On fourteen graphs, the \ktree{} takes $2.3$ to $27$ times less space than the CSR of cuSPARSE and, in geometric mean, beats it on left products on an H100 and an NVIDIA \GB{}, and on right products on the \GB{}; the H100 also lets it multiply a crawl of the web with 92 billion links that does not fit in device memory.
\end{abstract}



\maketitle

\section{Introduction}\label{sec:intro}

Iterative graph analytics are, in their inner loop, sparse matrix--vector products (SpMV): a step of breadth-first search (BFS) is a Boolean product, a step of Bellman--Ford the same product over the tropical (min,$+$) semiring, and an iteration of PageRank the plus-times product~\cite{graphblas}\cite[\S7.3]{golub-van-loan-matrix-book}. On GPUs, this product is bound by memory: bandwidth sets its speed, and capacity determines which graphs fit at all. Numerical linear algebra has thus long worked to move fewer bits per value or index~\cite{aliaga2021balanced,ginkgo_memory_accessor,frsz2,g-ANS}. Following compressed linear algebra~\cite{cla_vldbj,cla_cacm}, computation-friendly matrix formats~\cite{aware,bware,tuple_oriented_compression,mmrepair,louza2026hybrid_spmv,greener,francisco2022friendly,accelerating_gnn,unified_lossless_sparse,efficient_marino,rpq-k2tree-vldbj,arroyuelo2026boosting,compr-mm-bicliques,c5} run such products directly on compressed representations, in time proportional to their compressed size, which general grammar compressors cannot guarantee even for inner products~\cite{impossibility_gra_linalg}; thus, Boldi--Vigna-style encodings of adjacency lists, by gaps between successors and references to similar lists~\cite{webgraph,webgraph_rust,zuckerli}, grammar-compressed graphs and matrices~\cite{claude2010fast,tabei-grammar-compressed-pls,mmrepair} and wavelet trees~\cite[\S6.2]{navarro_book} have reached the GPU~\cite{gpu_traversal_compressed_graphs,c5,franzreb2025gpuwt}; Navarro surveys this line of work~\cite{friendly-compr-linalg-navarro}.

Standard sparse formats pay for every non-zero: the coordinate format (COO) stores both coordinates, the compressed sparse row format (CSR) stores a column index plus one pointer per row, and library implementations of both also store a value, even when every value is one. A CSR serves products by rows, so the \emph{left} product $x^{\top}\!A$ needs a transposed kernel, which scatters its outputs, or a second, column-oriented copy (CSC). The \ktree{}~\cite{k2tree-spire}\cite[\S9.2.1]{navarro_book}\cite[\S15.3]{ferragina_pearls} takes another route for Boolean matrices: each node splits its submatrix into $k^2$ blocks and keeps one bit per block, set iff the block holds a non-zero. Empty regions cost nothing, and the position of a non-zero is implicit in its path from the root, so \emph{neither values nor coordinates are stored}, and rows and columns are served alike. Introduced for web graphs more than fifteen years ago~\cite{k2tree-spire,k2tree-infosys}, the \ktree{} has since represented ternary relations, set operations and spatial data~\cite{interleaved-k2tree,k2tree-setops,set-operations-binary,compressed-quadtrees,k2-raster}, and supported sparse Boolean algebra: regular path queries (RPQs) in graph databases~\cite{rpq-k2tree-vldbj,arroyuelo2026boosting}, sums and products in cache-friendlier layouts~\cite{carmona-depth-first-2025,navarro-cache-friendly-2026}, and PageRank, where \cite{greener} shows it balances time, space and energy between grammar compression and Zuckerli~\cite{zuckerli}, a Boldi--Vigna-style compressor with entropy coding. To the best of our knowledge, it has never been implemented on a GPU.

The obstacle is the layout that makes the tree compact. Nodes are stored without pointers, level by level, and a node's children are found by $\mathrm{rank}$, the count of set bits up to a position in a bitvector~\cite{Munro96}\cite[\S4.2]{navarro_book}. $\mathrm{Rank}$ is cheap in isolation, also on GPUs~\cite{franzreb2025gpuwt}, but a top-down descent issues one lookup per visited node, each at an address known only after the previous one. This chain of dependent, irregular loads can be neither coalesced nor overlapped, and a GPU, which hides latency only behind many independent accesses, spends the descent waiting on memory.

\paragraph{Our approach} We traverse the tree level by level: the nodes of a level form a \emph{frontier} that a kernel expands in parallel into the next. Our key observation is that this traversal is the same for every input vector, so what $\mathrm{rank}$ provides (where the children of each node lie) is static. Our \emph{tree-only} variant keeps only the tree between products and recomputes the child offsets at every product, by an exclusive prefix sum (scan) over the frontier~\cite{BlellochTR90} (faster for batches) or from a \emph{directory} of offsets sampled every $256$ nodes, at about $0.1$ bits per node (faster for single vectors). The frontier (and the scanned offsets) is the only memory a product needs beyond the matrix and the vectors (its \emph{scratch}), and descending in \emph{tiles}, depth-first over contiguous ranges of each level, bounds it by the depth of the tree rather than its size. When memory is plentiful, the Morton codes (coordinates) of the last internal level can be stored with the matrix, at eight bytes per node: with this \emph{cached frontier}, a single-vector traversal is one kernel over the leaves. The same traversal computes the left product by swapping two coordinates at the leaves, and serves batches and any semiring with atomic addition.

None of these choices depends on the vector, so the device memory of a product is known after construction, before any kernel runs, and is the same on every GPU.

\paragraph{Contributions}
\begin{itemize}
\item \emph{A level-synchronous \ktree{} SpMV for the GPU} (\Cref{sec:algo}), whose child offsets, obtained by $\mathrm{rank}$ in a descent, are static data, scanned at each product or read from a sampled directory, and whose memory beyond the matrix and the vectors is bounded by the depth of the tree.
\item \emph{Right and left products from one structure} (\Cref{sec:leftbatch}): unlike CSR and Boldi--Vigna-style compressors, which serve one direction, the tree computes $Ax$ and $x^{\top}\!A$ from the same data, with the same memory up to the dense $x$ and $y$, for one vector or a batch, over the plus-times, Boolean and tropical semirings.
\item \emph{An evaluation on fourteen graphs of up to $1.2\!\times\!10^{9}$ edges} (\Cref{sec:eval}) on an NVIDIA H100, an L40S and a \GB{}: Wikidata relations, a subgraph of the Software Heritage (SWH) archive and six web and social graphs, against the cuSPARSE CSR~\cite{nvidia2026cusparse}, with its single-precision values, a recently proposed grammar-compressed GPU engine~\cite{c5}, \emph{bCSR} (a Boolean CSR without values that we implement as a simple baseline), and a multicore CPU descent of the same tree, with an ablation and a placement against measured hardware ceilings.
\item \emph{Two graphs beyond device memory} (\Cref{sec:capacity}): a $9.2\!\times\!10^{10}$-edge web crawl and a $2.8\!\times\!10^{10}$-edge SWH graph, whose trees are built out of core (by an external sort on disk) and multiplied, also within PageRank, on an H100 where neither CSR can be allocated.
\end{itemize}

\noindent
On an H100, the tree multiplies the crawl (whose cuSPARSE CSR would take 1.1~TB) in 37~GB of device memory and 0.66~s: $7.85\times$ faster than bCSR and $17.81\times$ faster than cuSPARSE at their best placements in host memory, and PageRank runs $5.26\times$ faster than any CSR with 32-bit indices can be streamed over PCIe~5.0. Where the CSR fits, the tree trades time for bytes: tree-only, it holds the fewest device bytes of any engine on the five largest graphs, 1.4~GB on the SWH subgraph of $1.2\!\times\!10^{9}$ edges against 10.3~GB for cuSPARSE, at $1.7$--$3.2\times$ the time of cuSPARSE, while on the left, among engines with one copy of the matrix, the tree is the fastest on ten graphs of fourteen. The \ktree{} is thus the format for when memory, not time, is the constraint.

\section{Background}\label{sec:prelim}

\subsection{The \ktree{}}
\begin{figure*}[t]
\centering
\begin{minipage}[c]{0.22\textwidth}\centering
\includegraphics[width=0.8\linewidth]{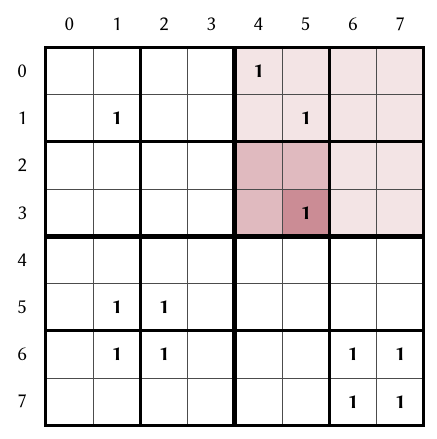}\par\smallskip
\includegraphics{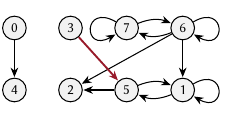}
\end{minipage}\hfill
\begin{minipage}[c]{0.76\textwidth}\centering
\includegraphics[width=\linewidth]{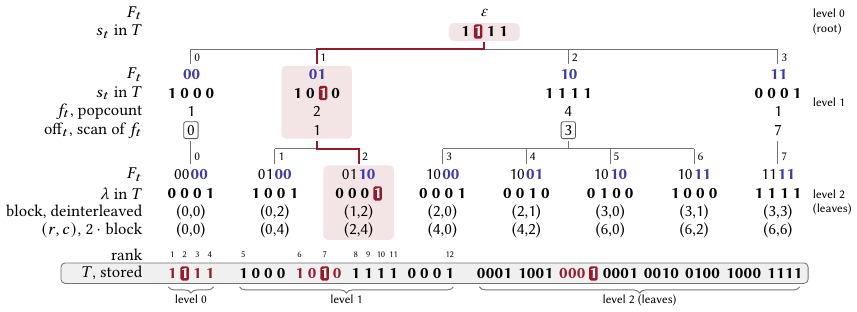}
\end{minipage}\par\smallskip
\includegraphics{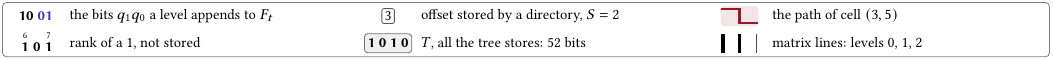}
\caption{A Boolean matrix (left), the directed graph it is the adjacency matrix of (below it), and its \ktree{} (right) as the level-order bitstring $T$, one level per block, each node above the range of its children. Each level shows what the traversal of \Cref{sec:algo} computes, in the notation of \Cref{alg:spmv}: the Morton code $F_t$ of node $t$, its signature $s_t$ (a leaf's is $\lambda$), its fan-out $f_t$, and the offset $\text{off}_t$ of its first child, which a descent obtains by $\mathrm{rank}$; at the leaves, the code is deinterleaved into the $2\times2$ block, doubled into its origin $(r,c)$. The path of cell $(3,5)$ runs through quadrant $q=\bin{01}$ of the root, $q=\bin{10}$ of the node of code $\bin{01}$ and $q=\bin{11}$ of leaf~$2$; its code $F_t=\bin{01\,10\,11}$ has row bits (odd positions) $\bin{011}=3$ and column bits (even positions) $\bin{101}=5$, and it is the edge $3\to5$ of the graph. At each node of the path, the bit taken is inverted in the tree and in $T$, where the three nodes are in red. Only $T$ is stored; the codes, fan-outs and offsets drawn above it are computed during a product.}\label{fig:k2tree}
\Description{A small Boolean matrix partitioned recursively into quadrants, with below it the directed graph of eight nodes and twelve edges whose adjacency matrix it is (an edge from i to j for each 1 in row i, column j; the edge from 3 to 5 is highlighted), next to its k2-tree written as the level-order bitstring of the tree, one level per block, each named on its right, with each node placed above its children and annotated with its Morton code, its number of children, and the offset of its first child; the eight leaves carry four-bit codes, the row and column of the two-by-two block they describe in the grid of such blocks, and that row and column doubled, the cell at its top-left corner. At the bottom, a light grey box shows the same 52 bits as the single stored bitvector, grouped by level, with no pointers, and above each 1 of the two internal levels its rank, from 1 to 12. A boxed legend below explains the marks: the two bits a level appends to a code, in color; the offsets stored by a directory, boxed; the path of cell (3,5), shaded; the ranks, not stored; the stored bitvector T; and the three line weights of the matrix, one per level.}
\end{figure*}

Let $A\in\{0,1\}^{n\times n}$ with $n=2^{\ell}$ (a rectangular matrix is padded with empty rows and columns, which cost nothing). The \ktree{} with $k=2$~\cite{k2tree-spire}\cite[\S9.2.1]{navarro_book}\cite[\S15.3]{ferragina_pearls} is the tree of the recursive $2\times2$ decomposition of $A$: the root has at least a child for each non-empty quadrant, and the recursion stops at single cells. Each internal node stores a \emph{signature} $s$ of $k^2=4$ bits: with the quadrants numbered $q=\bin{q_1q_0}=2q_1+q_0\in\{0,\dots,k^2-1\}$, where the bit $q_1$ selects the lower half of the rows and the bit $q_0$ the right half of the columns, bit $q$ of $s$ is set iff quadrant $q$ is non-empty. We write numbers in binary as $\bin{\cdot}$, most significant bit first, and signatures as bitstrings in the order of $T$, bit $0$ first. In \Cref{fig:k2tree}, the root has $s=\mathsf{1111}$, and its child $q=\bin{01}=1$, the top-right quadrant, has $s=\mathsf{1010}$, since only its quadrants $\bin{00}=0$ and $\bin{10}=2$ hold non-zeros. Concatenating the signatures in level order yields the bitstring $T$, which is all the tree stores, on disk as on the device: no pointers and no coordinates, with levels implicit (bottom of \Cref{fig:k2tree}). The children of the node whose bit is at offset $p$ of $T$ start at offset $k^2\cdot\mathrm{rank}_1(T,p)$, where $\mathrm{rank}_1(T,p)$ counts the set bits of $T[0..p]$: every step down the tree is a $\mathrm{rank}$ query~\cite{rpq-k2tree-vldbj}\cite[\S9.2.1]{navarro_book}. A matrix with $m$ non-zeros induces $\bigoh(m\log(n^2/m))$ nodes; our trees take from $2.4$ bits per edge on a web crawl to $28$ on a citation relation (\Cref{tab:graphs}).

\subsection{Semiring products}
A semiring $\mathbb{S}=(D,\oplus,\otimes,\bar 0,\bar 1)$ is a set $D$ with two operations such that $(D,\oplus,\bar 0)$ is a commutative monoid, $(D,\otimes,\bar 1)$ a monoid, $\otimes$ distributes over $\oplus$, and $\bar 0$ annihilates under $\otimes$. It defines the product $y=Ax$ by $y_i=\bigoplus_{j}A_{ij}\otimes x_j$. Our $A$ is the Boolean adjacency matrix of a graph without edge weights: an absent edge is $\bar 0$ and every edge carries the same weight $w$, so $y_i=\bigoplus_{j:\,A_{ij}=1}w\otimes x_j$. Contributions reach $y_i$ from many threads in no fixed order, which the associativity and commutativity of $\oplus$ make harmless (in floating point, up to rounding), so our kernel only requires an atomic $\oplus$ on the GPU (\Cref{tab:semirings}; $\max$ and $\min$ use integer atomics, as non-negative floats order as their bits read as integers). \textbf{Plus-times}, with $w=1$, is the kernel of PageRank and of any power iteration on an unweighted graph~\cite[\S7.3]{golub-van-loan-matrix-book}; \textbf{Boolean} $(\vee,\wedge)$ is a step of BFS~\cite[\S22.2]{cormen-book-edition3}; \textbf{tropical} $(\min,+)$, with $w=1$, is a hop-count relaxation round of Bellman--Ford~\cite[\S24.1]{cormen-book-edition3}, as generalized to any semiring by GraphBLAS~\cite{graphblas}. All three are cross-checked (\Cref{sec:correct}); timings use plus-times. Only the matrix is Boolean: $x$ and $y$ hold arbitrary values of $D$ (FP32 here).

\begin{table}[t]
\centering
\caption{The three semirings we implement, and the bit-sliced form of the Boolean one (\Cref{sec:leftbatch}): domain, operations, identities, storage of a vector entry, and the contribution $w\otimes x_j$ of a non-zero, applied by one GPU atomic ($\infty$ is $10^9$).}\label{tab:semirings}
\footnotesize
\setlength{\tabcolsep}{2.5pt}
\begin{tabular}{llccccccl}
\toprule
semiring & $D$ & $\oplus$ & $\otimes$ & $\bar0$ & $\bar1$ & entry & leaf & atomic\\
\midrule
plus-times & $\mathbb{R}$             & $+$    & $\times$ & $0$      & $1$ & FP32 & $x_j$        & \texttt{atomicAdd}\\
Boolean    & $\{0,1\}$                & $\vee$ & $\wedge$ & $0$      & $1$ & FP32 & $[x_j\neq0]$ & \texttt{atomicMax}\\
\quad bit-sliced & $\{0,1\}^{32}$     & $\vee$ & $\wedge$ & $0^{32}$ & $1^{32}$ & 1 bit & $x_j$ & \texttt{atomicOr}\\
tropical   & $\mathbb{Z}_{\ge0}\cup\{\infty\}$ & $\min$ & $+$ & $\infty$ & $0$ & FP32 & $1+x_j$      & \texttt{atomicMin}\\
\bottomrule
\end{tabular}
\end{table}

\section{Level-Synchronous \ktree{} SpMV}\label{sec:algo}

A GPU rewards independent work and coalesced accesses and punishes chains of dependent, irregular loads such as a top-down descent. We thus propose to descend the tree one level at a time (\Cref{alg:spmv}): the nodes of a level form a frontier (in level order), and one kernel expands all of them into the next level, with independent lookups issued by adjacent threads at adjacent positions.

\begin{algorithm}[t]
\caption{Level-synchronous \ktree{} SpMV $y=Ax$ over a semiring $\mathbb{S}$. Node $t$ of the frontier has Morton code $F_t$, signature $s_t$, fan-out $f_t=\popc(s_t)$, and first child $\text{off}_t$ in the next level; each \textbf{parallel for} gives one thread to each node.}\label{alg:spmv}
\begin{algorithmic}[1]
\Require tree $T$, one signature (nibble) per node in level order, with levels $0$ (root) to $\ell-1$ (leaves); vector $x$; semiring $\mathbb{S}$ with edge weight $w$; frontier $F$, the Morton codes of the nodes of a cached level $d_0$
\State $y \gets \bar0^{\,n}$ \Comment{fill kernel}
\For{$d = d_0$ \textbf{to} $\ell-2$} \Comment{one level per step}
  \State $\text{off}_t \gets \sum_{u<t} f_u$ for every node $t$ of $F$ \label{alg:off} \Comment{\textsc{ChildOffset}}
  \If{$d < \ell-2$} \Comment{expand to the next frontier}
    \ParFor{each node $t$ of $F$}
      \For{the $j$-th set bit of $s_t$, at position $q$}
        \State $F'_{\text{off}_t+j} \gets (F_t\ll 2)\,|\,q$ \label{alg:expand} \Comment{code of child $j$}
      \EndFor
    \EndParFor
    \State $F \gets F'$ \Comment{two buffers, alternated}
  \Else \Comment{last internal level: leaves visited in place}
    \ParFor{each node $t$ of $F$}
      \For{the $j$-th set bit of $s_t$, at position $q$}
        \State $\lambda \gets$ signature of leaf $\text{off}_t+j$ \label{alg:leaf}
        \State $(r,c) \gets 2\cdot\textsc{Deinterleave}\big((F_t\ll 2)\,|\,q\big)$ \label{alg:deint}
        \For{each set bit of $\lambda$, at position $2\,dr+dc$}
          \State $y_{r+dr} \gets y_{r+dr}\oplus\big(w\otimes x_{c+dc}\big)$ \label{alg:atomic} \Comment{atomic}
        \EndFor
      \EndFor
    \EndParFor
  \EndIf
\EndFor
\State \Return $y$
\end{algorithmic}
\end{algorithm}

In \Cref{alg:spmv}, levels are numbered from $0$ (root), to $\ell-1$ (leaves: each a $2\times2$ block of cells); within a level, nodes are numbered from $0$ in level order. The \emph{Morton code} $F_t$ of node $t$ interleaves the row and column bits of its block, two bits per level of its path, and its \emph{signature} $s_t$ is its nibble in $T$. Level order lists the children of consecutive nodes consecutively. Hence, those of node $t$ occupy positions $\text{off}_t$ to $\text{off}_t+f_t-1$ of level $d+1$, where $\text{off}_t=\sum_{u<t}f_u$; computing these offsets is the step $\textsc{ChildOffset}$ (line~\ref{alg:off}), by a scan or from a directory (\Cref{sec:offsets}), and it replaces the $\mathrm{rank}$ queries of a descent. Child $j$ of node $t$ is the quadrant $q$ of the $j$-th set bit of $s_t$ (line~\ref{alg:expand}). The leaves are never written to a frontier: the thread of a node of the last internal level reads the signature $\lambda$ of each of its leaves (line~\ref{alg:leaf}), decodes the origin $(r,c)$ of its block (line~\ref{alg:deint}), and adds each non-zero cell to $y$ by one atomic of the semiring (line~\ref{alg:atomic}). The loop starts from a level $d_0$ whose codes are computed once and kept: levels $0$ to $d_0-1$ are expanded on the host (\Cref{sec:tiles}); with the frontier cached, $d_0=\ell-2$ and only the \textbf{else} branch runs.

\subsection{Frontier, Morton codes and leaves}\label{sec:frontier}
A frontier entry is a $64$-bit Morton code $F_t$, i.e., the quadrants on the path from the root, two bits per level. Descending to quadrant $q=\bin{q_1q_0}$ appends $q_1$ and $q_0$, $(F_t\ll2)\,|\,q$ (line~\ref{alg:expand}): one shift and one OR on one word (where a coordinate pair would need two updates) and the frontier stays one coalesced array. At the leaves, $F_t$ is decoded once. Deinterleaving a code of $d$ quadrants (the row concatenates the $q_1$ bits, the column the $q_0$ bits) gives a block in the grid of blocks of side $n/2^{d}$: a leaf's code gives its $2\times2$ block, whose origin, the top-left cell, is twice that, i.e., a $0$ appended to each coordinate. In \Cref{fig:k2tree}, the leaf of code $F_t=\bin{01\,10}$ has row bits $\bin{01}=1$ and column bits $\bin{10}=2$: it is the block $(1,2)$, of origin $(\bin{010},\bin{100})=(2,4)$, and its only set cell, $(3,5)$, contributes $x_5$ to $y_3$. Signatures are stored one per nibble, so reading a node is a byte load, a shift and a mask. The leaves read no indices and no values, which a valued format loads for every non-zero.

\subsection{Child offsets: a scan or a sampled directory}\label{sec:offsets}
A level-synchronous traversal visits the nodes of a level in left-to-right order, which is exactly the order $\mathrm{rank}$ counts. In \Cref{fig:k2tree}, the four nodes of level~1 have signatures $s_t=\mathsf{1000}$, $\mathsf{1010}$, $\mathsf{1111}$, $\mathsf{0001}$ and fan-outs $f_t=1,2,4,1$. Their scan, $\text{off}_t=0,1,3,7$, gives each node $t$ the offset of its first child, and their total ($8$) is the size of the next frontier: each counts the $1$s of the level before its node. One scan per level thus replaces every $\mathrm{rank}$ of the level with no structure beyond the tree; it is the data-parallel form of the sequential traversal that lets a CPU SpMV over the whole tree do without $\mathrm{rank}$~\cite{greener}, and the way GPU BFS places the output of a frontier~\cite{scalable_gpu_graph_traversal,high_performance_gpu_graph_traversal}. We compute it with CUB's single-pass scan with decoupled look-back~\cite{merrill2016singlepass}.

The offsets depend only on the tree, so they are the same at every product and can be computed once, at construction, and sampled. One $32$-bit offset per node would cost eight times the tree, so the \emph{directory} keeps one every $S$ nodes: for each block of $S$ consecutive nodes, the offset of the first child of its first node. $S$ trades space for time. The offset of any node is the sample of its block plus the popcount of the signatures that precede it in the block, i.e., of at most $S/16$ $64$-bit words of the tree. In \Cref{fig:k2tree}, with $S=2$, the directory stores the boxed offsets $0$ and $3$, and the node of code $\bin{01}$ finds its first child at $0+\popc(\mathsf{1000})=1$. This one-level sampled rank directory~\cite[\S4.2.1]{navarro_book} is lighter than the two-level structures that answer arbitrary $\mathrm{rank}$ queries on GPUs~\cite{franzreb2025gpuwt}: it costs $0.06$--$0.11$ bits per node on our graphs, and a level becomes one kernel, with no scan. It is faster for single vectors; in a batch, the scan is paid once while the directory lookup is repeated per vector, and the scan wins (\Cref{sec:where}).

\subsection{Threads and work, on the running example}\label{sec:mapping}
Every kernel gives one thread to each node of its frontier; a batched product of $B$ vectors gives each node $B$ threads, one per vector, which follow the same path through the tree and differ only in the entries of $x$ and $y$ they read and update. In \Cref{fig:k2tree}, $\ell=3$ and the root is expanded on the host, so $d_0=1$: a product starts from the codes $\bin{00}$, $\bin{01}$, $\bin{10}$, $\bin{11}$ of level~1 and traverses in one kernel of four threads, one per non-empty $4\times4$ block, running the \textbf{else} branch of \Cref{alg:spmv}. The fan-outs $1,2,4,1$ scan to the offsets $0,1,3,7$ (line~\ref{alg:off}), so the thread of code $\bin{10}$ owns leaves $3$ to $6$. Leaf~$5$, of code $\bin{10\,10}$ and origin $(6,0)$, holds the single cell $(6,1)$ and adds $x_1$ to $y_6$. The four threads issue $1$, $3$, $4$ and $4$ updates, which collide: $y_1$ receives cell $(1,1)$ from the thread of code $\bin{00}$ and $(1,5)$ from that of $\bin{01}$; hence the need for the semiring's atomic. In general, the threads of the last kernel are the non-empty $4\times4$ blocks, at most $\min(m,n^2/16)$ for $m$ non-zeros, and from $0.9$ to $296$ million on our graphs. The tree-only engine launches one tile of them at a time, $2^{22}$ threads in blocks of $256 = 2^8$, at least $15$ waves of resident blocks on the H100 and $19$ on the L40S ($14$ on the \GB{}, with tiles of $2^{20}$), so a partly filled last wave idles little.

\emph{Load balance.} The work of a thread is bounded by the arity of the tree, not by the degrees of the graph: an internal node writes at most $k^2=4$ codes, and a node of the last internal level issues $1$ to $k^4=16$ updates. A CSR kernel with one thread per row inherits the degree distribution, so a hub row stalls its warp; the remedies are a warp per row~\cite{2009_bell_garland_spmv}, merge-based splits~\cite{2016_merrill_garland_spmv}, balanced layouts~\cite{aliaga2021balanced,csr5}, or rows sorted by length. The \ktree{}, instead, splits a hub row across one thread per non-empty $4\times4$ block, so the imbalance within a warp is at most $k^4 = 16$ to one on any graph, and with the loops over a signature unrolled, its threads diverge only in how many updates they issue; we do not sort the frontier by work, which would break the contiguity of children.

\emph{Memory access.} Adjacent threads read adjacent codes, signatures, offsets and leaves; only the accesses to $x$ and $y$ are irregular, and consecutive codes are consecutive in Z-order, so on a clustered matrix the rows and columns a warp touches share cache lines.

\emph{Leaf aggregation.} The cells of a $4\times4$ block write to at most four entries of $y$, one per row of the block (or per column, if on the left), so its thread merges their contributions in registers by $\oplus$ and issues at most four atomics instead of sixteen. For single vectors, the threads of a warp whose $4\times4$ blocks lie in the same four rows of $A$ (columns, on the left) find each other by one warp-wide match, sum their values by register shuffles, and only the lowest lane issues the atomics. This warp step removes the collisions of left products, where many rows hit a column of high in-degree; on right products it pays on the \GB{} and the L40S but not on the H100, so, like the tile, we set it per GPU (\Cref{sec:archs}).

\emph{Granularity.} A thread per leaf would write the leaves, the widest level, as a frontier for the next kernel to read; a thread per $8\times8$ block takes $1.1$--$2.8\times$ the time on the seven Wikidata relations and \texttt{popular-3k-python}, as its work bound grows fourfold.

\subsection{Bounded scratch: leaves in place, a host-side top, and tiles}\label{sec:tiles}
Besides the matrix and the vectors, a product needs the \emph{scratch}: the device memory it allocates for the Morton codes of the frontier and, when the offsets come from a scan rather than the directory, for the offsets. Three properties of the level-order layout keep it small. First, the scan reads the fan-outs straight from the tree, so no fan-out array is stored. Second, levels never shrink with depth, since every node has a child, so the leaves are the widest level; but the leaves of a node of the last internal level are contiguous, so its kernel visits them in place (the \textbf{else} branch of \Cref{alg:spmv}) and the widest frontier is never written. Third, a kernel launch costs more than the work of the first, tiny levels, so we expand them once on the host and start every product from $d_0$, the deepest level with at most $2^{16}$ nodes ($0.52$~MB of codes).

\begin{figure}[t]
\centering
\includegraphics{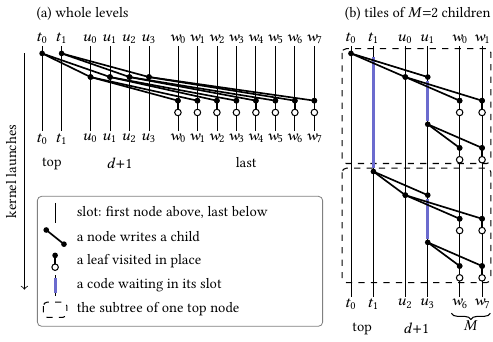}
\caption{The descent of a small tree as a data-flow network, drawn like a prefix circuit~\cite{par_prefix_comp1980}: (a) whole levels take 3 launches, but each buffer is as wide as its level; (b) tiles of at most $M=2$ children take 10 launches, one tile at a time and depth-first, but only $M$ slots per level.}\label{fig:tiles}
\Description{Two panels side by side, of vertical lines, the slots of the frontier buffers grouped by level (top, next level, last internal level), with time running downwards, one row per kernel launch on an axis shared by both panels; each line is labeled above and below with the first and the last node its slot holds. In panel (a), three launches expand whole levels: two top nodes write four children, which write eight, whose leaves are then visited; the last group is eight slots wide, and each slot holds one node. Below it, a boxed legend repeats each mark with its meaning. In panel (b), with tiles of two children, each level has two slots, and ten launches descend the same tree depth-first, one tile at a time, so the lower-level slots are reused: the labels change from top to bottom. Dashed boxes enclose the launches of each top node's subtree, and thick violet lines mark the codes that wait in their slots between launches.}
\end{figure}

The remaining scratch is as wide as the last internal levels and grows with the matrix (\Cref{fig:tiles}a); tiles remove this dependence (\Cref{fig:tiles}b). In level order, the children of a contiguous range of nodes form a contiguous range so that a level can be cut into \emph{tiles}, ranges of nodes with at most $M$ children, each descended to the leaves before the next is expanded: a depth-first traversal over ranges of nodes. One buffer of at most $M$ codes per level suffices, so the scratch is $\bigoh(M\ell)$, bounded by the depth of the tree rather than its size: at most $8M$ bytes of codes per level below the cached top ($33.6$~MB with $M=2^{22}$), plus, with the scan, $4M$ bytes of offsets and the temporary storage of the scan. Each tile costs one launch, so the tile size trades scratch for launches. The host plans the tiles, which depend only on the tree, and allocates the exact scratch once, before any product; for single vectors, it records a product's launches once as a CUDA graph and resubmits it with a single call. We use $M=2^{22}$ ($2^{20}$ on the \GB{}, \Cref{sec:archs}); matrices whose levels are narrower than $M$ are traversed as without tiles.

\subsection{Tree-only and cached frontier}\label{sec:dial}
Every level of codes kept with the matrix saves its expansion at every product and costs device memory; we evaluate the two extremes. In the \emph{tree-only} configuration, \TO{}, a matrix keeps the tree, its directory and the codes of level $d_0$, and the scratch is transient: one scratch can serve every matrix on the device, as suits a knowledge graph whose thousands of relations are thousands of matrices~\cite[\S2.1.1]{hogan_kg_book}. In the \emph{cached-frontier} configuration, \CF{}, the Morton codes of the last internal level are computed when the matrix is loaded and kept with it, at eight bytes per node of that level. A product then expands no internal level: each thread takes its code from the cached array and its leaf offset from the directory, so a single-vector traversal is one kernel over the leaves.

\subsection{Left products and batches}\label{sec:leftbatch}
The \ktree{} treats rows and columns alike, so the left product $y^{\top}=x^{\top}\!A$ needs no other structure: the kernel reads the same tree, frontier and scratch, and only swaps the decoded coordinates at the leaf, so that cell $(r,c)$ adds $x_r$ to $y_c$. Its memory differs from that of a right product only in the dense $x$ and $y$, whose lengths swap for a rectangular matrix. A CSR stores for each row its non-zero columns (the successors of a node) and so do Boldi--Vigna-style compressors~\cite{webgraph,webgraph_rust,zuckerli}, also on GPUs~\cite{gpu_traversal_compressed_graphs}. The left product reads such a structure by rows but accumulates by columns: a transposed kernel gives row $i$ to a thread, which adds $x_i$ to $y_j$ for every column $j$ of the row, so threads whose rows share a column update the same $y_j$ atomically and serialize when they collide; the alternative is a second, CSC copy. Both directions are needed for direction-optimizing traversals~\cite{beamer2013direction} and for RPQs, which follow edges and inverse edges~\cite{LA-RPQ,rpq-k2tree-vldbj}. A batch of $B$ vectors is an $n\times B$ row-major array, so a leaf visit updates $B$ contiguous outputs from $B$ adjacent threads: the traversal, scans, and reads of the tree are paid once per batch, and only the leaf updates grow with $B$. Batched products take their offsets from the scan.

\emph{Bit-sliced Boolean batches.} Over the Boolean semiring, an entry needs one bit: as multi-source BFS does on CPUs~\cite{merrier_pvldb2014}, we also pack $32$ vectors into every $32$-bit entry: a batch of $B$ vectors is an $n\times B/32$ array of words, and a leaf ORs a word of $x$ into $y$ by one \texttt{atomicOr}, skipped when it would set no new bit of $y$. The traversal and the scratch are those of any batch of $B/32$ entries; only the semiring changes, and with it the vectors shrink $32$-fold.

\section{Experimental Setup}\label{sec:setup}

\paragraph{Platform and protocol}\label{sec:hw} All experiments run on an NVIDIA H100 of the GRICAD Kraken cluster\footnote{\url{https://gricad.univ-grenoble-alpes.fr/en/news/2026/inauguration-kraken/}} (\texttt{sm\_90}, 99.9~GB, PCIe~5.0 to the host), with CUDA~13.1 (driver 610.57.04) and \texttt{nvcc -O3}; \Cref{sec:archs} repeats them on an NVIDIA \GB{} and an L40S. We place ourselves in the ``compress once, multiply many times'' setting: a matrix is built once on the host, timed separately (\Cref{sec:space}), and multiplied by many vectors not known in advance, each product timed on the device between two CUDA events. Times are medians over ten runs of $50$ products for our engine and over five runs of each baseline binary; the interquartile range stays within $0.3\%$ of the median for us and $2.4\%$ for the baselines. Batches have $B=128$ vectors, or $16$ on \texttt{popular-3k-python} and \texttt{it-2004}, whose dense blocks $X,Y\in\mathbb{R}^{n\times128}$ would take $47$ and $42$~GB, too much to run every engine next to its matrix on every GPU of \Cref{sec:archs}. Dense vectors are FP32 in every engine. Energy per product is read from the NVML energy counter over a loop of at least 1.5~s of products, as in~\cite{c5}; following~\cite{hpc_benchmark}, the tables give absolute times. Unless stated otherwise, single-vector products use the directory ($S=256$), batched ones the scan, and tiles hold $2^{22}$ nodes ($2^{20}$ on the \GB{}, \Cref{sec:archs}). Sizes are decimal (1~MB $=10^6$ bytes, 1~GB $=10^9$ bytes).

\paragraph{Datasets}\label{sec:datasets} We use fourteen graphs in two groups (\Cref{tab:graphs}). The first holds Boolean relations, the data of knowledge graphs and their engines~\cite{google_kg,amazon_kg,cikm_pennino_2026,milleniumdb,rpq-k2tree-vldbj}. Seven are relations of the Wikidata knowledge graph~\cite{wikidata-cacm,the_most_out_wd}\cite[\S10.1.4]{hogan_kg_book}, in the version of~\cite{time-space-rpq} ($9.6\!\times\!10^{8}$ edges over $5{,}419$ relations), from the data package of~\cite{c5} (Zenodo \texttt{doi:\allowbreak 10.5281/\allowbreak zenodo.22677747}): six relations and the citation relation \texttt{wd\_cites\_work} ($1.7\!\times\!10^{8}$ edges). The eighth, \texttt{popular-3k-python}, from the same package, is a subgraph of the Merkle DAG~\cite{swh_dag_2019} of SWH, the universal archive of public source code~\cite{swh_cacm,di-cosmo-archiving,swh-ecosystems-book-chapter}, with $4.6\!\times\!10^{7}$ nodes and $1.2\!\times\!10^{9}$ edges, distributed in BVGraph~\cite{Boldi2020GraphCompression}, the format of the WebGraph framework of Boldi and Vigna~\cite{webgraph,webgraph_rust}. The second group holds six graphs of the Laboratory for Web Algorithmics (LAW), as distributed (BVGraphs in layered-label-propagation order~\cite{layered_label_prop}): four web crawls (\texttt{indochina-2004}, \texttt{uk-2002}, \texttt{arabic-2005}, \texttt{it-2004}: the data the \ktree{} was designed for) and two harder cases: a social network (\texttt{ljournal-2008}) and the links of the English Wikipedia (\texttt{enwiki-2013}). The package of~\cite{c5} ships RePair grammars for the Wikidata relations and \texttt{popular-3k-python}; we compress the LAW graphs with the RePair builder from~\cite{mmrepair}. Two more graphs (\texttt{gitlab-all} and \texttt{eu-2015}) serve only the tests beyond device memory (\Cref{sec:capacity}).

\paragraph{Baselines}\label{sec:baselines} For \textbf{cuSPARSE}~\cite{nvidia2026cusparse}, the standard, we time \texttt{SpMV}, transposed \texttt{SpMV} for the left product, and, for batches, the fastest of \texttt{SpMM} with its default algorithm, \texttt{CSR\_ALG2} and \texttt{CSR\_ALG3}; its released baseline in~\cite{c5} allocates managed as well as device memory, so we also time the same calls on device memory in our harness and report the faster. In \textbf{Ginkgo}~1.11~\cite{ginkgo}, a second sparse library, we screen every format (CSR with its four strategies, COO, ELL, Hybrid, SELL-P) on right, batched and left products, the left one on a CSC copy, as Ginkgo computes it, and time every exact format with our protocol. For PageRank (\Cref{sec:capacity}) we also run \textbf{cuGraph}~26.10~\cite{cugraph} with the same parameters and number of iterations as ours. The \textbf{grammar} engine~\cite{c5}\footnote{Zenodo \texttt{doi:\allowbreak 10.5281/\allowbreak zenodo.23139248}.} is a recently proposed level-synchronous GPU engine over a RePair~\cite{repair,repair_journal} grammar of the matrix; we run its released binaries, which implement right products only, and its grammar file, entropy-coded with asymmetric numeral systems (ANS)~\cite{mmrepair}, is the compressed baseline for space. \textbf{bCSR}, a Boolean CSR with row pointers and column indices but no values, is a simple baseline that we implement (not part of our proposal) to isolate what a CSR saves without values. On the right, we report the faster of the two CSR kernels of Bell and Garland~\cite{2009_bell_garland_spmv}, with one thread (scalar) or one warp (vector) per row; on the left, it adds with atomics, and a batch gives a thread to each row and vector. A \textbf{CPU} descent of the same tree and directory, by $\mathrm{rank}$, on the 48 host cores allotted to the GPU, shows what the GPU contributes.

\paragraph{Correctness}\label{sec:correct} cuSPARSE, a CPU CSR oracle and our engine recompute the $y$ of the grammar engine on its $x$ (the tropical one against a CPU min-plus product) and agree on the six relations, bit for bit on the Boolean and tropical semirings and to single precision on plus-times (relative difference below $10^{-6}$). On all fourteen graphs, with integer entries, both configurations match a CPU product bit for bit, right and left, for single vectors and batches of $B$ distinct vectors (offsets from the directory or the scan) over the three semirings and the bit-sliced one; bCSR and the CPU descent match it on plus-times.

\section{Results}\label{sec:eval}
The results answer five questions: how small the tree is (\Cref{sec:space}); what it costs in time and saves in device memory when every format fits on the device (\Cref{sec:time,sec:ws}); why (\Cref{sec:features,sec:where,sec:roofline}); how the answer changes with the GPU (\Cref{sec:archs}); and what the tree enables when the CSR does not fit (\Cref{sec:capacity}).

\begin{table*}[t]
\centering
\caption{The fourteen graphs (Wikidata relations; SWH subgraph; LAW web and social graphs): \ktree{} structure, persistent size of each format (gram.: RePair grammar; BV: BVGraph; cuSP.: cuSPARSE CSR) and construction time~\cite{mmrepair}.}\label{tab:graphs}
\footnotesize
% GENERATED by extract_results.py -- do not edit by hand.
\begin{tabular}{lr rrr rrrrrrr rr}
\toprule
& & \multicolumn{3}{c}{\ktree{}} & \multicolumn{7}{c}{persistent size (MB)} & \multicolumn{2}{c}{construction (s)}\\
\cmidrule(lr){3-5}\cmidrule(lr){6-12}\cmidrule(lr){13-14}
graph & nnz & nodes & levels & bits/edge & tree & \TO{} & \CF{} & gram. & BV & cuSP. & bCSR & \TO{} & RePair\\
\midrule
\texttt{wd\_sports\_team} & 1.14M & 5.82M & 19 & 20.5 & 2.91 & 3.25 & 11.5 & 1.93 & 2.61 & 10.4 & 5.87 & 0.389 & 0.380\\
\texttt{wd\_cast\_member} & 1.03M & 5.82M & 18 & 22.5 & 2.91 & 3.16 & 10.5 & 2.21 & 2.23 & 8.96 & 4.83 & 0.382 & 0.350\\
\texttt{wd\_citizenship} & 3.06M & 7.72M & 22 & 10.1 & 3.86 & 4.25 & 19.5 & 3.25 & 7.42 & 36.0 & 23.7 & 0.662 & 0.700\\
\texttt{wd\_occupation} & 4.60M & 17.29M & 22 & 15.0 & 8.65 & 9.06 & 39.2 & 5.38 & 13.1 & 50.6 & 32.2 & 1.34 & 1.16\\
\texttt{wd\_subclass\_of} & 2.02M & 5.71M & 21 & 11.3 & 2.85 & 3.22 & 12.1 & 3.19 & 3.39 & 22.1 & 14.0 & 0.437 & 0.450\\
\texttt{wd\_country} & 10.09M & 16.93M & 24 & 6.7 & 8.46 & 9.09 & 43.4 & 9.63 & 18.9 & 121 & 80.6 & 1.89 & 2.18\\
\texttt{wd\_cites\_work} & 166.68M & 1.18G & 24 & 28.3 & 590 & 606 & 1,784 & 459 & 421 & 1,362 & 695 & 83.5 & 95.3\\
\midrule
\texttt{popular-3k-python} & 1.22G & 1.27G & 26 & 4.2 & 633 & 644 & 3,009 & 490 & 378 & 9,931 & 5,057 & 228 & 630\\
\midrule
\texttt{ljournal-2008} & 79.02M & 281.82M & 23 & 14.3 & 141 & 145 & 533 & 180 & 109 & 654 & 338 & 24.5 & 44.7\\
\texttt{enwiki-2013} & 101.36M & 413.42M & 23 & 16.3 & 207 & 212 & 805 & 220 & 166 & 828 & 422 & 33.3 & 61.8\\
\texttt{indochina-2004} & 194.11M & 116.75M & 23 & 2.4 & 58.4 & 59.4 & 261 & 79.1 & 35.4 & 1,583 & 806 & 28.0 & 70.2\\
\texttt{uk-2002} & 298.11M & 232.62M & 25 & 3.1 & 116 & 118 & 520 & 203 & 84.6 & 2,459 & 1,267 & 47.3 & 128\\
\texttt{arabic-2005} & 640.00M & 439.70M & 25 & 2.7 & 220 & 223 & 988 & 302 & 147 & 5,211 & 2,651 & 100 & 279\\
\texttt{it-2004} & 1.15G & 795.23M & 26 & 2.8 & 398 & 404 & 1,786 & 555 & 261 & 9,371 & 4,768 & 187 & 509\\
\bottomrule
\end{tabular}

\end{table*}

\begin{table*}[t]
\centering
\caption{Per-vector time and working set on the fourteen graphs: right, left and batched products ($B{=}128$; $^{\dagger}$: $B{=}16$) of our engine (\TO{}: tree-only; \CF{}: cached frontier), the grammar engine (gram.), bCSR, cuSPARSE (cuSP.) and Ginkgo (gko., its fastest format; the left product on a CSC copy). Graph names are those of \Cref{tab:graphs}, shortened.}\label{tab:perf}
\footnotesize
\setlength{\tabcolsep}{3.3pt}
% GENERATED by extract_results.py -- do not edit by hand.
\begin{tabular}{l rrrrrr rrrrr rrrrr rrrrr}
\toprule
& \multicolumn{6}{c}{right product $Ax$ (ms)} & \multicolumn{5}{c}{left product $x^{\top}\!A$ (ms)} & \multicolumn{5}{c}{batched $AX$ (ms/vector)} & \multicolumn{5}{c}{working set (MB)}\\
\cmidrule(lr){2-7}\cmidrule(lr){8-12}\cmidrule(lr){13-17}\cmidrule(lr){18-22}
graph & \TO{} & \CF{} & gram. & bCSR & cuSP. & gko. & \TO{} & \CF{} & bCSR & cuSP. & gko. & \CF{} & gram. & bCSR & cuSP. & gko. & \TO{} & \CF{} & gram. & bCSR & cuSP.\\
\midrule
\texttt{sports\_team} & 0.074 & 0.027 & 0.063 & 0.008 & 0.016 & 0.011 & 0.081 & 0.035 & 0.046 & 0.052 & 0.020 & 0.010 & 0.011 & 0.003 & 0.002 & 0.002 & 21.2 & 12.9 & 9.33 & 7.32 & 11.9\\
\texttt{cast\_member} & 0.075 & 0.024 & 0.068 & 0.022 & 0.017 & 0.011 & 0.080 & 0.029 & 0.019 & 0.024 & 0.018 & 0.009 & 0.011 & 0.002 & 0.002 & 0.002 & 19.1 & 11.8 & 11.1 & 6.10 & 10.2\\
\texttt{citizenship} & 0.095 & 0.051 & 0.098 & 0.013 & 0.032 & 0.017 & 0.372 & 0.383 & 1.29 & 1.29 & 0.040 & 0.025 & 0.058 & 0.019 & 0.011 & 0.011 & 43.0 & 31.0 & 34.6 & 35.3 & 47.5\\
\texttt{occupation} & 0.199 & 0.091 & 0.129 & 0.026 & 0.042 & 0.036 & 0.583 & 0.419 & 1.15 & 1.04 & 0.053 & 0.043 & 0.071 & 0.024 & 0.014 & 0.014 & 79.5 & 53.1 & 43.4 & 46.1 & 64.5\\
\texttt{subclass\_of} & 0.069 & 0.029 & 0.076 & 0.009 & 0.021 & 0.013 & 0.098 & 0.059 & 0.962 & 0.961 & 0.028 & 0.015 & 0.032 & 0.011 & 0.006 & 0.006 & 24.8 & 18.4 & 19.7 & 20.3 & 28.4\\
\texttt{country} & 0.195 & 0.116 & 0.283 & 0.070 & 0.119 & 0.074 & 0.647 & 0.499 & 6.87 & 6.09 & 0.117 & 0.074 & 0.201 & 0.065 & 0.040 & 0.040 & 105 & 83.6 & 121 & 121 & 161\\
\texttt{cites\_work} & 13.3 & 3.08 & 3.39 & 1.94 & 1.57 & 1.59 & 13.9 & 3.58 & 2.63 & 3.03 & 1.13 & 1.36 & 1.35 & 0.309 & 0.325 & 0.339 & 1,063 & 1,861 & 1,478 & 772 & 1,439\\
\midrule
\texttt{popular-3k} & 16.8 & 11.0 & 5.95 & 4.17 & 5.30 & 5.32 & 15.2 & 9.64 & 19.9 & 16.9 & 5.32 & 3.78$^{\dagger}$ & 3.08$^{\dagger}$ & 1.17$^{\dagger}$ & 2.33$^{\dagger}$ & 2.08$^{\dagger}$ & 1,365 & 3,374 & 2,793 & 5,422 & 10,296\\
\midrule
\texttt{ljournal} & 3.17 & 1.16 & 1.51 & 0.513 & 0.405 & 0.410 & 3.30 & 1.33 & 1.21 & 0.869 & 0.410 & 0.471 & 0.561 & 0.095 & 0.080 & 0.082 & 523 & 576 & 626 & 380 & 697\\
\texttt{enwiki} & 4.64 & 1.68 & 1.61 & 0.745 & 0.475 & 0.483 & 5.18 & 2.26 & 2.49 & 1.81 & 0.490 & 0.690 & 0.599 & 0.093 & 0.078 & 0.080 & 572 & 839 & 715 & 456 & 861\\
\texttt{indochina} & 1.48 & 1.03 & 1.42 & 1.41 & 0.844 & 0.852 & 1.30 & 0.861 & 5.13 & 2.71 & 0.845 & 0.319 & 0.461 & 0.141 & 0.105 & 0.108 & 247 & 321 & 430 & 865 & 1,642\\
\texttt{uk-2002} & 2.38 & 1.38 & 2.50 & 0.944 & 1.41 & 1.42 & 2.54 & 1.59 & 6.08 & 3.16 & 1.41 & 0.615 & 1.14 & 0.279 & 0.179 & 0.182 & 462 & 668 & 1,050 & 1,415 & 2,607\\
\texttt{arabic} & 4.74 & 2.99 & 3.31 & 1.97 & 2.79 & 2.80 & 5.18 & 3.45 & 19.8 & 12.1 & 2.79 & 1.16 & 1.51 & 0.444 & 0.321 & 0.325 & 632 & 1,170 & 1,471 & 2,833 & 5,393\\
\texttt{it-2004} & 8.57 & 5.30 & 6.11 & 3.55 & 5.01 & 5.02 & 9.54 & 6.35 & 34.0 & 20.9 & 5.06 & 2.34$^{\dagger}$ & 2.94$^{\dagger}$ & 1.03$^{\dagger}$ & 1.91$^{\dagger}$ & 1.90$^{\dagger}$ & 1,000 & 2,116 & 2,595 & 5,098 & 9,701\\
\bottomrule
\end{tabular}

\end{table*}

\subsection{Space and construction}\label{sec:space}\label{sec:construction}
\begin{figure}[t]
\centering
\includegraphics[width=\linewidth]{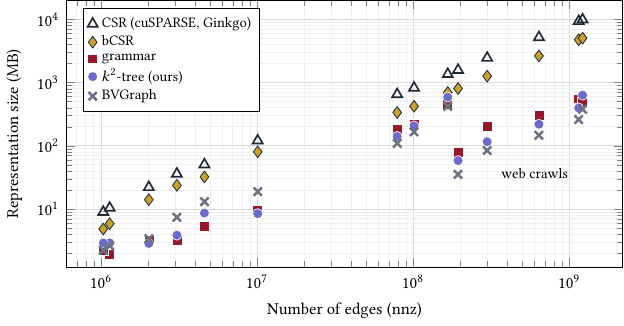}
\caption{Representation size vs.\ number of edges (log--log). The \ktree{} stays below the CSR of cuSPARSE and Ginkgo and bCSR throughout, over tenfold on the web crawls; BVGraph is smaller on ten graphs, larger on four relations.}\label{fig:space}
\Description{Log-log plot of representation size against edge count on fourteen graphs: the k2-tree, the grammar, bCSR, the cuSPARSE CSR, and the WebGraph BVGraph. The k2-tree lies below both the cuSPARSE CSR and bCSR on every graph, by more than ten times on four web crawls; the BVGraph lies below the k2-tree on ten graphs and above it on four Wikidata relations.}
\end{figure}

The tree is $2.3$--$27\times$ smaller than the cuSPARSE CSR and $1.2$--$14\times$ smaller than bCSR, most on the web crawls ($21$--$27\times$ and $11$--$14\times$), as its size follows the clustering of the non-zeros rather than their number (\Cref{tab:graphs,fig:space}): from $2.4$ bits per edge on a web crawl to $28$ on the citation relation. Persistent sizes count what stays on the device between products: for our engine, the tree, the directory and the codes of level $d_0$ (\TO{}), plus the cached frontier (\CF{}). Packed into $\lceil\log_2 n\rceil$ bits, the column indices of a web crawl would still take $8$--$10\times$ the tree. The tree is smaller than the RePair grammar by up to $1.75\times$ on the six LAW graphs (third block of \Cref{tab:graphs}) and on \texttt{wd\_subclass\_of} and \texttt{wd\_country}; on the other six graphs of the first two blocks, whose rows repeat most, the grammar is $0.62$--$0.84\times$ the tree. BVGraph (built by us for the Wikidata relations, with the parameters of the LAW graphs) is $0.60$--$0.90\times$ the tree on ten graphs, but $1.19$--$2.23\times$ on the four relations with $1.0$--$1.4$ non-zeros per row, where it pays a degree and a first gap per row; it serves successors only. The tree-only engine adds $1.5$--$13\%$ to the tree. The cached frontier, eight bytes per node of the last internal level, makes \CF{} $2.9$--$4.8\times$ \TO{}, the price of its speed, yet \CF{} stays below the cuSPARSE CSR on eleven of the fourteen graphs, down to $0.17$--$0.21\times$ on the web crawls, and exceeds it only on the two smallest relations and \texttt{wd\_cites\_work} ($1.11$--$1.31\times$).

The host builds the tree on a single core, from the parsed edge list to the tree-only engine on the device, in $0.38$--$1.9$~s on the six relations and at most $228$~s on \texttt{popular-3k-python}, the time of $5{,}100$--$22{,}000$ single-vector products; caching the frontier adds $0.9$--$9.3\%$. RePair takes $0.86$--$1.2\times$ as long on the Wikidata relations and $1.8$--$2.8\times$ on the other seven graphs.

\subsection{Right, left and batched products}\label{sec:time}
\emph{Right product.} \CF{} takes $0.98$--$2.2\times$ the time of cuSPARSE on the six relations and $0.98$--$3.5\times$ over the fourteen graphs, on par on \texttt{wd\_country} and \texttt{uk-2002} and furthest on the social graphs (\Cref{tab:perf,sec:features}). Tree-only, re-deriving the frontier at every product costs $1.6$--$9.8\times$ the time of cuSPARSE, and the gap to \CF{} follows the nodes per non-zero: $1.4$--$1.7\times$ on the web crawls, with fewer than one, and $4.3\times$ on the citation relation, with seven. \CF{} is $1.1$--$2.8\times$ faster than the grammar engine on twelve graphs, including all seven Wikidata relations, and $1.05$--$1.85\times$ slower on \texttt{enwiki-2013} and on \texttt{popular-3k-python}, whose repeated rows a grammar evaluates once. The CPU descent of the tree on 48 cores takes $24$--$114\times$ the time of the tree-only GPU engine. bCSR, which moves half the bytes of cuSPARSE, takes $0.41$--$1.67\times$ its time and is the fastest engine on nine graphs, cuSPARSE on four and Ginkgo on \texttt{wd\_cast\_member}; \CF{} beats bCSR on \texttt{indochina-2004} ($0.73\times$).

\emph{Left product.} From the same tree and scratch, \CF{} computes $x^{\top}\!A$ in $0.061$--$1.21\times$ the time of cuSPARSE's transposed kernel on the Wikidata relations, faster on five of seven, and in $0.28$--$0.57\times$ on the web crawls and \texttt{popular-3k-python}. The left product turns the gather of a right product into a scatter, whose updates collide on relations with few columns (\Cref{sec:features}). It slows cuSPARSE by $1.4$--$51\times$ over its own right product, ours by at most $7.6\times$, and bCSR, which scatters with atomic additions, by up to $108\times$: among engines with one copy of the matrix, \CF{} is the fastest on the left on nine graphs and tree-only on one, bCSR and cuSPARSE on two each, and \CF{} beats bCSR on eleven. A CSR recovers its right-product speed on the left only with a second, CSC copy, at twice the memory.

\begin{figure}[t]
\centering
\includegraphics[width=\linewidth]{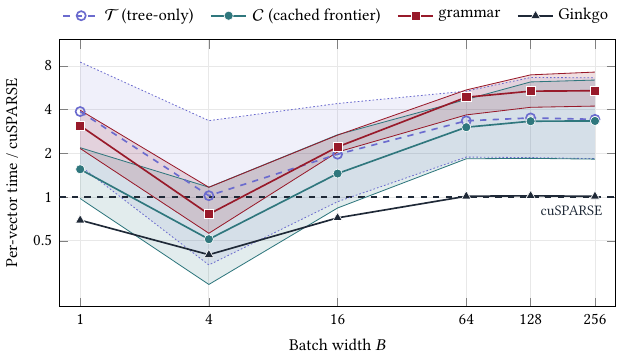}
\caption{Per-vector time against batch width, relative to cuSPARSE at the same width (its \texttt{SpMV} at $B=1$, its fastest \texttt{SpMM} above): geometric mean (line) and range (band) over the seven Wikidata relations (log--log); Ginkgo is its fastest exact format at each width.}\label{fig:batched}
\Description{Log-log plot of per-vector time relative to cuSPARSE against batch width from 1 to 256, with geometric means over seven graphs as lines and their ranges as bands, for the tree-only k2-tree, the k2-tree with cached frontier and the grammar engine, and as a line alone for Ginkgo's fastest exact format. The cached frontier is below cuSPARSE at width 4 and about three times its time from 64 on; the tree-only line falls towards it as the width grows; the grammar engine lies above both from 16 on; Ginkgo is below cuSPARSE up to 16 and level with it from 64.}
\end{figure}

\emph{Batches.} Batching amortizes the traversal, which is where the two configurations differ: at $B=128$ the tree-only engine is within $12\%$ of \CF{}, which takes $1.2$--$8.9\times$ the time of cuSPARSE's best \texttt{SpMM} and is $1.1$--$2.7\times$ faster than the grammar engine on eleven graphs. cuSPARSE is the fastest engine in batch on eleven graphs and bCSR, with one thread per row and vector, on the other three, as neither has a traversal to pay. Across batch widths on the Wikidata relations (\Cref{fig:batched}), \CF{}, already a single-kernel traversal, gains less than cuSPARSE's \texttt{SpMM}: it takes $1.56\times$ the time of cuSPARSE at $B=1$ and $0.51\times$ at $B=4$ (geometric means), and settles at about three times its time from $B=64$. The tree-only engine gains most, $5.1\times$ per vector from $B=1$ to $B=128$ (geometric mean). Bit-sliced, a Boolean batch costs little more than one product: on \texttt{popular-3k-python}, tree-only, $64$ vectors take $21$~ms on the right and on the left, $1.7\times$ a single Boolean product and $11.7\times$ less than the FP32 batch of the same vectors (0.73~GB of dense blocks against 23.4~GB).

\emph{Ginkgo.} Ginkgo's fastest exact format takes $0.84\times$ the time of cuSPARSE on the right product (geometric mean, range $0.52$--$1.02\times$), faster on the six relations. In batch, it matches cuSPARSE's \texttt{SpMM} ($1.01\times$, range $0.89$--$1.04\times$), ahead of it up to $B=16$ on the Wikidata relations and level from $B=64$ (\Cref{fig:batched}). On the left, it multiplies a CSC copy in $0.18\times$ the time of cuSPARSE's transposed kernel: the speed of a right product, paid with the matrix twice on the device. \CF{} takes $1.89\times$ the time of Ginkgo's best on the right ($0.97\times$ on \texttt{uk-2002}), $3.26\times$ in batch and $2.46\times$ on the left, serving both directions from one tree.

\emph{BFS.} A multi-source breadth-first search iterates the batched Boolean product to its fixed point (on \texttt{popular-3k-python}, for $32$ levels), along or against the edges, marking newly reached entries in dense $n\times B$ tables. On \texttt{wd\_subclass\_of}, \texttt{wd\_cites\_work} and \texttt{popular-3k-python}, it yields the same distances as a CPU search and cuSPARSE \texttt{SpMM} with a CSR oriented for each direction, in $0.99$--$3.91\times$ its time tree-only and $0.89$--$3.60\times$ with the frontier cached, slowest on the citation relation, with seven nodes per non-zero. On the \GB{}, where bandwidth is scarce (\Cref{sec:archs}), the factors shrink to $1.06$--$1.63\times$ and $0.91$--$1.46\times$. Tree-only, the search on \texttt{popular-3k-python} uses 8.3~GB on the device against 17.6~GB.

\subsection{Working set}\label{sec:ws}
\begin{figure}[t]
\centering
\includegraphics[width=\linewidth]{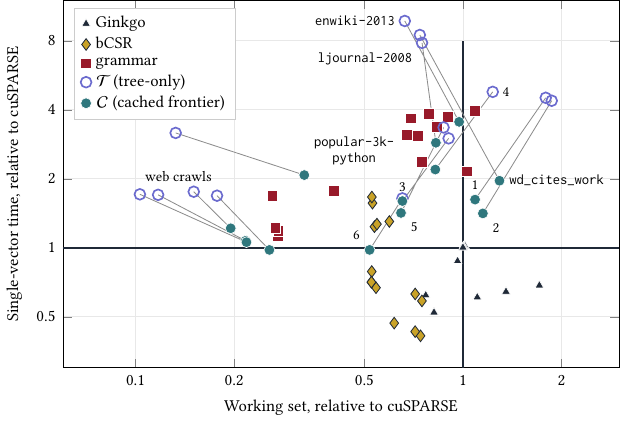}
\caption{Working set during a single right product (matrix $+$ dense vectors $+$ scratch) against its time, both relative to cuSPARSE (log scales); a line joins the two configurations of a graph; Ginkgo is its fastest format. Points 1--6 are the six relations in the row order of \Cref{tab:graphs}; the others are named.}\label{fig:tradeoff}
\Description{Log-log scatter plot of working set against single-vector time, both relative to cuSPARSE, for the tree-only k2-tree, the k2-tree with cached frontier, the grammar engine and bCSR on fourteen graphs. Most k2-tree points lie left of the cuSPARSE line; the tree-only points of four web crawls and of the popular-3k-python graph lie about ten times to its left and two to three times above it, and the bCSR points cluster just below and to the left of the reference.}
\end{figure}

Tree-only, our engine holds less device memory than cuSPARSE during a product on eleven graphs of fourteen, and the least of any engine on the five largest. This \emph{working set} counts the matrix, the dense vectors and, for our engine, the scratch; \Cref{fig:tradeoff} plots it against time, both relative to cuSPARSE. The tree-only engine needs $0.10$--$0.18\times$ on the web crawls and \texttt{popular-3k-python}, $0.66$--$0.75\times$ on \texttt{ljournal-2008}, \texttt{enwiki-2013} and the citation relation, and $0.65$--$0.90\times$ on three other relations. On \texttt{popular-3k-python} it holds 1.4~GB against 10.3~GB for cuSPARSE and $0.49\times$ the working set of the grammar engine. The cached frontier needs $0.20$--$1.3\times$ the working set of cuSPARSE. On \texttt{wd\_sports\_team}, \texttt{wd\_cast\_member} and \texttt{wd\_occupation}, the tree-only engine holds $1.2$--$1.9\times$ the bytes of cuSPARSE, as the scratch of its tiles weighs more than a tree of a few megabytes; bCSR holds $0.32$--$0.82\times$ the working set of the tree-only engine on every Wikidata relation but \texttt{wd\_country}, and on \texttt{ljournal-2008} and \texttt{enwiki-2013}. On \texttt{popular-3k-python} and the four web crawls, the tree-only engine holds $3.1$--$5.1\times$ fewer bytes than bCSR: it is the low-memory end of the trade-off in \Cref{fig:tradeoff}, at $1.7$--$3.2\times$ the time of cuSPARSE, and the cached frontier spends memory to move along it.

\subsection{Which graphs favor the tree}\label{sec:features}
Two graph properties explain where the tree wins and loses. The first is the \emph{clustering} of the non-zeros, which sets the tree's work per update: the nodes per non-zero (\Cref{tab:graphs}) and the non-zeros per non-empty $4\times4$ block, i.e., per thread of the last kernel. On the web crawls and \texttt{popular-3k-python}, a block holds $4.1$--$7.7$ non-zeros and the tree $0.60$--$1.04$ nodes per non-zero, and a tree-only product takes $1.4$--$1.6\times$ the time of its leaves. On the two smallest relations and the citation relation, a block holds $1.1$ non-zeros and the tree $5.1$--$7.1$ nodes per non-zero, so every update pays for its own path: the tree is only $2.3$--$3.6\times$ smaller than the cuSPARSE CSR, and tree-only takes $4.4$--$8.5\times$ its time.

The contrast is sharpest between the web crawls and the two social graphs, whose long rows (\texttt{popular-3k-python} and the LAW graphs hold $15$--$28$ non-zeros per row) cuSPARSE streams alike, at $0.84$--$0.93\times$ the copy bandwidth (\Cref{sec:roofline}). On the web crawls, with $0.60$--$0.78$ nodes per non-zero, the tree reads $0.31$--$0.40$ bytes of matrix per non-zero, its leaves update $180$--$201$~G non-zeros per second, and \CF{} takes $0.98$--$1.22\times$ the time of cuSPARSE. On the two social graphs, with $3.6$--$4.1$ nodes per non-zero, it reads $1.8$--$2.1$ bytes, its leaves reach $53$--$61$~G/s, and \CF{} takes $2.9$--$3.5\times$.

The second property, for the left product, is the \emph{distribution of non-zeros across columns}. \texttt{wd\_citizenship}, \texttt{wd\_occupation} and \texttt{wd\_country} hold $430$--$5{,}800$ non-zeros per column on $1{,}747$--$10{,}610$ columns: their updates collide on few entries of $y$, and the transposed kernel of cuSPARSE takes $25$--$51\times$ the time of its own right product, the tree, which merges colliding updates before its atomics (\Cref{sec:mapping}), $4.3$--$7.6\times$: $0.08$--$0.40\times$ the time of cuSPARSE. Where the non-zeros spread over many columns (\texttt{wd\_sports\_team}, \texttt{wd\_cast\_member}, \texttt{ljournal-2008}, the citation relation), the tree, which treats rows and columns alike, takes $1.14$--$1.30\times$ its right-product time, and the transposed kernel of cuSPARSE $1.39$--$3.19\times$.

\subsection{Where the time goes}\label{sec:where}
Together, the techniques of \Cref{sec:algo} make a single-vector product $5.8$--$15\times$ faster on the six relations, and those that only bound memory cost at most $7\%$. Starting from a level-synchronous descent that scans a separate array of fan-outs, we add them one at a time, in the order below. A preallocated CUB scan fed straight from the tree gains $1.6$--$1.9\times$; visiting the leaves in place saves their frontier at $0.93$--$1.04\times$ the time, and expanding the top levels on the host gains $1.3$--$1.7\times$; tiles of $2^{22}$ nodes cost at most $7\%$ ($1.00$--$1.07\times$ the time); the directory gains $1.2$--$1.5\times$ and the CUDA graph $1.05$--$1.11\times$, completing the \TO{} configuration; the cached frontier gains $1.5$--$2.5\times$, and leaf aggregation in registers $1.2$--$1.4\times$. In all, a single-vector product drops from $0.36$--$0.67$~ms to $0.024$--$0.12$~ms.

On the seven Wikidata relations and \texttt{popular-3k-python}, the directory beats the scan on every single-vector product tree-only, by $1.17$--$1.58\times$; with the frontier cached, the scan takes $0.94$--$1.29\times$ the time of the directory; in batch, the scan takes $0.58$--$0.84\times$ the time of the directory, hence our split. On a single vector, the block size of the directory changes the time by at most $11\%$ between $S=64$ and $256$. Tree-only, shrinking the tile from whole levels to $2^{14}$ nodes cuts the scratch $14$--$1{,}830\times$ and makes it independent of the matrix; tiles of $2^{22}$ nodes take $0.99$--$1.24\times$ the untiled time (at most $1.04\times$ on the six relations), and $2^{22}$ is the fastest tile on every graph. Below it, the fixed cost of a tile takes over: $1.0$--$1.9\times$ the untiled time at $2^{20}$ nodes, $20$--$61\times$ at $2^{14}$.

\subsection{Hardware efficiency}\label{sec:roofline}
The CSR engines are bound by the bytes of their matrix, the tree by the updates of its vectors (\Cref{tab:roofline}). We time every kernel of a product with Nsight Systems and compare each engine with two ceilings that we measure: the copy bandwidth, 2{,}154~GB/s, and the rate of vector updates $y_o \mathrel{+}= x_i$, which a microbenchmark puts at $175$~G/s for consecutive entries with atomics, $172$~G/s with plain stores, and $6.9$~G/s for scattered entries with atomics. We report the eight largest graphs, whose CSR far exceeds the caches.

Counting only what CSR engines must move (the matrix, $x$ read once, $y$ read and written once), cuSPARSE streams $1{,}818$--$2{,}005$~GB/s on seven graphs, $0.84$--$0.93\times$ the copy bandwidth, and bCSR $614$--$1{,}579$~GB/s; on the citation relation, whose scattered reads of $x$ this count leaves out, $939$ and $402$~GB/s. Their time follows the matrix bytes: eight per non-zero for cuSPARSE, four for bCSR.

The tree reads $0.31$--$0.53$ bytes of matrix per non-zero on the web crawls and \texttt{popular-3k-python} ($1.35$--$2.47$ with the frontier cached, whose codes it reads instead of the internal levels), and its kernels keep the GPU busy for at least $99.7\%$ of a product ($92.3\%$ on the smallest relations). With the frontier cached, its leaves sustain $190$--$221$~G non-zeros per second on the web crawls, $1.09$--$1.26\times$ the rate of one consecutive atomic per non-zero, as leaf aggregation (\Cref{sec:mapping}) issues fewer atomics than non-zeros; on \texttt{popular-3k-python}, $111$~G/s, and on the social graphs and the citation relation, whose non-zeros cluster less, $54$--$68$~G/s. A CSR accumulates a row in a register, whereas the tree updates $y$ for each row of a block: up to twelve bytes of vector traffic per non-zero, against a fraction of that for the matrix. Tree-only, expanding the internal levels takes a further $26$--$37\%$ of a product on the web crawls and \texttt{popular-3k-python}, and $59$--$74\%$ on the social graphs and the citation relation, which have more nodes per non-zero. The compressed matrix has thus left the critical path: a faster tree needs fewer bytes per update, not per node, such as larger leaves whose row segments accumulate in registers, or narrower vector entries.

\begin{table}[t]
\caption{Time for a single right product on the eight largest graphs, against the ceilings measured on the H100: the share of the internal levels in a tree-only product, the rate of the leaves, and the bandwidth of the CSR engines.}\label{tab:roofline}
\footnotesize
\setlength{\tabcolsep}{4pt}
% GENERATED by extract_results.py -- do not edit by hand.
\begin{tabular}{l r rr rr}
\toprule
& levels & \multicolumn{2}{c}{leaves (G nnz/s)} & \multicolumn{2}{c}{CSR (GB/s)}\\
\cmidrule(lr){3-4}\cmidrule(lr){5-6}
graph & (\% of \TO{}) & \TO{} & \CF{} & bCSR & cuSP.\\
\midrule
\texttt{wd\_cites\_work} & 74 & 46.6 & 53.7 & 402 & 939\\
\midrule
\texttt{popular-3k-python} & 30 & 104 & 111 & 1343 & 2002\\
\midrule
\texttt{ljournal-2008} & 60 & 61.5 & 68.0 & 776 & 1818\\
\texttt{enwiki-2013} & 59 & 52.9 & 59.2 & 633 & 1879\\
\texttt{indochina-2004} & 26 & 180 & 190 & 614 & 2005\\
\texttt{uk-2002} & 37 & 201 & 221 & 1579 & 1938\\
\texttt{arabic-2005} & 32 & 201 & 216 & 1482 & 1990\\
\texttt{it-2004} & 32 & 201 & 220 & 1487 & 1995\\
\midrule
measured ceiling & & \multicolumn{2}{c}{175} & \multicolumn{2}{c}{2154}\\
\bottomrule
\end{tabular}

\end{table}

\subsection{Across GPU architectures}\label{sec:archs}
Where memory bandwidth is scarce, the bytes the tree saves buy time as well as capacity: on an NVIDIA \GB{}, which copies at 241~GB/s against 2{,}154~GB/s on the H100, \CF{} takes $0.79\times$ the time of cuSPARSE on the right and $0.39\times$ on the left (geometric means), against $1.58\times$ and $0.43\times$ on the H100. We repeat the measurements on the \GB{} (\texttt{sm\_121}, 128~GB shared with the CPU, CUDA~13.0, driver 580.178.04) and an NVIDIA L40S (\texttt{sm\_89}, 48~GB, CUDA~12.6, driver 555.42.02), with the same engine, inputs, protocol and batch widths (\Cref{tab:archs}); every cross-check that runs passes again. From the H100 to the \GB{}, a single-vector right product takes $6.8\times$ longer with cuSPARSE (geometric mean), $6.1\times$ with bCSR, $5.2\times$ with the grammar engine, and $3.4\times$ with our engine with the frontier cached, $3.9\times$ tree-only, so our factor to cuSPARSE halves on the right and drops from $3.29\times$ to $2.09\times$ in batch. On the \GB{}, \CF{} is faster than cuSPARSE on seven graphs on the right and on twelve on the left, and, among engines with one copy of the matrix, the fastest on the left on twelve. The L40S lies in between: cuSPARSE takes $1.67\times$ its H100 time and \CF{} $1.09\times$, bringing \CF{} to $1.03\times$ the time of cuSPARSE on the right, $0.39\times$ on the left and $2.14\times$ in batch. Of the launch-dependent settings, the split of offsets carries over (the directory for single vectors, the scan for batches), the tile size does not: $2^{22}$ nodes on the H100 and the L40S, $2^{20}$ on the \GB{}, which replays a CUDA-graph node in $0.43$~$\mu$s against $0.79$ (\Cref{tab:archs}). The warp step of leaf aggregation (\Cref{sec:mapping}) runs on left products only on the H100, in both directions on the \GB{} and the L40S.

\begin{table}[t]
\centering
\caption{The engine on the H100, the L40S and the \GB{}, by decreasing memory bandwidth: device and launch costs; the time of each engine relative to cuSPARSE on the same GPU (geometric mean over the fourteen graphs, range in brackets; below $1$: faster than cuSPARSE); the tile size, which depends on the cost of a launch. On the L40S, the batched row covers thirteen graphs: cuSPARSE cannot run the \texttt{arabic-2005} batch alongside its matrix in 48~GB.}\label{tab:archs}
\footnotesize
\setlength{\tabcolsep}{3pt}
% GENERATED by extract_results.py -- do not edit by hand.
\begin{tabular}{lccc}
\toprule
 & H100 & L40S & \GB{}\\
\midrule
compute capability & 9.0 & 8.9 & 12.1\\
SMs & 132 & 142 & 48\\
memory (GB) & 100 & 48 & 128\\
L2 cache (MiB) & 60 & 96 & 24\\
copy bandwidth (GB/s) & 2154 & 673 & 241\\
kernel launch ($\mu$s) & 1.90 & 1.71 & 2.04\\
graph node ($\mu$s) & 0.79 & 0.57 & 0.43\\
\midrule
\multicolumn{4}{l}{\emph{time relative to cuSPARSE}} \\
right, \TO{} & 3.40 (1.64--9.76) & 2.20 (0.535--8.59) & 1.96 (0.616--5.37)\\
right, \CF{} & 1.58 (0.979--3.54) & 1.03 (0.339--3.29) & 0.794 (0.339--1.65)\\
right, grammar & 2.36 (1.12--3.97) & 1.83 (0.748--3.56) & 1.80 (0.984--2.98)\\
left, \CF{} & 0.434 (0.061--1.53) & 0.390 (0.059--1.55) & 0.388 (0.109--1.22)\\
batched, \CF{} & 3.29 (1.22--8.86) & 2.14 (1.13--3.69) & 2.09 (1.17--3.39)\\
\midrule
tile (nodes) & $2^{22}$ & $2^{22}$ & $2^{20}$\\
\bottomrule
\end{tabular}

\end{table}

\subsection{Beyond device memory}\label{sec:capacity}
\begin{figure}[t]
\centering
\includegraphics[width=\linewidth]{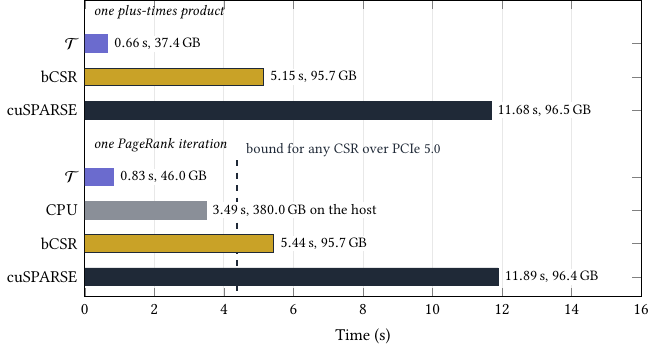}
\caption{Beyond device memory on the H100: one plus-times product and one PageRank iteration on \texttt{eu-2015} ($9.2\!\times\!10^{10}$ edges), with the tree-only engine (\TO{}), bCSR and cuSPARSE streamed from host memory at their best placement, and the 48 host cores; each bar gives the time and is labeled with the memory held. Dashed: the least time in which any CSR of the predecessors, with 32-bit indices and without values, crosses PCIe~5.0 at every iteration.}\label{fig:capacity}
\Description{Horizontal bar chart on a logarithmic time axis, in two groups. One plus-times product: the k2-tree in 0.66 seconds holding 37.4 GB of device memory, bCSR in 5.15 seconds holding 95.7 GB, cuSPARSE in 11.68 seconds holding 96.5 GB. One PageRank iteration: the k2-tree in 0.83 seconds with 46.0 GB, the CPU in 3.49 seconds over a CSR of 380.0 GB in host memory, bCSR in 5.44 seconds and cuSPARSE in 11.89 seconds, both holding about 96 GB; a dashed vertical line at about 4.4 seconds marks the lower bound for any CSR with 32-bit indices streamed over PCIe 5.0, to the right of the k2-tree bar.}
\end{figure}
Beyond device memory, the tree holds 37~GB and leaves the host link idle, whereas any CSR must cross it at every product: on the web crawl \texttt{eu-2015}, the tree is thus $7.85\times$ faster than bCSR and $17.81\times$ faster than cuSPARSE (\Cref{fig:capacity}), and on the less clustered \texttt{gitlab-all}, $3.71\times$ faster than cuSPARSE at $1.08\times$ the time of bCSR. We first take \texttt{gitlab-all}, the SWH export of all of GitLab.com (December 2020)~\cite{swh_dag_2019}: a BVGraph like \texttt{popular-3k-python}, with $23\times$ its edges ($2.8\!\times\!10^{10}$) over $1.1\!\times\!10^{9}$ nodes. On the empty H100 (99.9~GB), neither CSR can be allocated: bCSR needs 129.0~GB with $x$ and $y$, and cuSPARSE, with 64-bit indices and offsets, 352.4~GB. The tree takes 27.1~GB, a tree-only product 37.4~GB with $x$ and $y$; its four deepest levels exceed $2^{32}$ nodes, so offsets take 64 bits.

A plus-times product takes 0.70~s, a Boolean one 0.59~s, and a batch of four 0.34~s per vector (medians of three runs of ten), each equal to a host product over the edge list. We stream the smallest CSR cuSPARSE accepts, row panels of at most $2^{28}$ non-zeros with 32-bit indices (227.7~GB with its values), in four placements, each checked against the edge list: in managed memory, a product takes 36.4~s; with the leading panels on the device and the rest read in place, 4.55~s; from pinned host memory (zero-copy, as in EMOGI~\cite{emogi}), 7.44~s. The fastest streams the panels that do not fit through two device buffers, each copy overlapping the product of the previous panel, at 56~GB/s: 2.60~s. bCSR keeps 83.8~GB of its 116.0~GB on the device and streams the other 32.2~GB at every product: 0.65~s, filling 95.4~GB of the device, $2.55\times$ the tree, and the host link. The tree is built out of core: the nodes of a level are the distinct prefixes of the sorted Morton codes, so the host sorts them by distribution~\cite[\S5.3]{ferragina_pearls} on local disk and emits every level in one pass: 940~s (551~s reading network storage) and 110~GB of host memory.

The clustering decides which end of the trade-off wins: \texttt{eu-2015}, a LAW crawl of the web domains of the European Union, has $3.3\times$ the edges of \texttt{gitlab-all} ($9.2\!\times\!10^{10}$) over as many nodes, yet a tree of as many nodes ($5.5\!\times\!10^{10}$, $0.60$ per edge against $1.9$): 27.5~GB, 2.4 bits per edge. bCSR would need 384.3~GB, nearly all of four such GPUs, and cuSPARSE 1118.6~GB; the tree-only product holds 37.4~GB, as on \texttt{gitlab-all}, and takes 0.66~s, a Boolean one 0.47~s and a batch of four 0.31~s per vector, all equal to the host product. bCSR keeps 83.7~GB on the device and moves 287.7~GB over the host link per product in 5.15~s, while cuSPARSE moves 657.0~GB in 11.68~s. The tree is built in 2{,}526~s with 115~GB of host memory.

\textbf{PageRank.} With damping $d=0.85$, i.e., teleport probability $1-d=0.15$~\cite{brin1998anatomy,pagerank}\cite[\S21.2]{ir_book}, we run $100$ iterations of
\[ r_{t+1}=\tfrac{1-d}{n}\mathbf{1}+d\,\big(A^{\top}D^{-1}r_t+\tfrac{s_t}{n}\mathbf{1}\big),\qquad r_0=\tfrac{1}{n}\mathbf{1}, \]
where $D$ holds the out-degrees (one right product) and $s_t$ the rank of the nodes without out-links; an iteration is one left product, $A^{\top}(D^{-1}r_t)$, and two small kernels. On \texttt{gitlab-all}, an iteration takes $0.81$~s, with $0.76$~s in the left product. The same iterations on the $48$ host cores allotted to the GPU, over a CSR of the predecessors in host memory, take $1.81$~s each, $2.24\times$ longer, with rank vectors within an $\ell_1$ distance of $10^{-6}$ and the same top $100$ nodes. cuSPARSE, streaming a CSR of the predecessors as above, takes $2.78$~s per iteration, $3.46\times$ ours, and cuGraph~\cite{cugraph} cannot load the graph, as its cuDF edge table holds fewer than $2^{31}$ rows. bCSR, streaming 40.8~GB of its CSR of the predecessors at every iteration, takes $0.93$~s ($1.16\times$ ours) while holding 95.5~GB on the device, compared with our 46.1~GB. On \texttt{eu-2015}, an iteration takes $0.83$~s in 46.0~GB: $4.19\times$ faster than the 48 host cores over a CSR of 380.0~GB ($3.49$~s), $6.54\times$ faster than bCSR, which streams 296.3~GB per iteration ($5.44$~s), and $14.30\times$ faster than cuSPARSE ($11.89$~s). No engine that runs these iterations over a CSR with 32-bit indices can match this: as every node is active at every iteration of the power method, active-edge transfers~\cite{subway,ascetic,HyTGraph}, unified memory~\cite{grus,HALO} and zero-copy access~\cite{emogi} alike must bring the part of the CSR that the device cannot hold, which, without values and with the device filled, is 280.1~GB, at least $4.38$~s at the 64~GB/s of PCIe~5.0, $5.26\times$ our iteration. Where the CSR fits, the trade-off reverses: on \texttt{popular-3k-python}, the tree takes $17.1$~ms per iteration against $81.3$~ms on the host, but cuSPARSE takes $6.2$~ms and cuGraph $13.1$~ms in 43.9~GB of device memory against 1.74~GB. Counting construction from the local disk ($169$~s for the CSR of the predecessors of \texttt{gitlab-all}, $789$~s for the tree), the tree is repaid after about $620$ iterations; on \texttt{eu-2015}, it builds faster than the CSR ($2{,}204$ against $2{,}467$~s).
\section{Limitations}\label{sec:limitations}
\begin{enumerate}
\item \emph{The tree needs clustering}, hence a good order of the nodes (we keep the order of each distribution). On the citation relation, the tree is only $2.3\times$ smaller than the cuSPARSE CSR (the grammar $3.0\times$), and the cached frontier takes $1.96\times$ the time of cuSPARSE while holding $1.29\times$ its bytes.
\item \emph{Compressors that exploit repetition are smaller.} BVGraph is $0.60$--$0.90\times$ the tree on ten graphs, the grammar smaller on six; yet the tree serves both products from one structure.
\item \emph{Time where bandwidth is plentiful.} On the H100, a CSR engine is the fastest on the right and in batch on every graph, and on the eight largest \CF{} takes $0.98$--$3.5\times$ the time of cuSPARSE and $0.98$--$3.0\times$ its energy, as the leaves update $y$ once per row of a block; the tree buys memory with that time, and time on the \GB{} (\Cref{sec:archs}).
\item \emph{Beyond device memory, the trade-off follows the clustering.} On \texttt{gitlab-all} (7.8 bits per edge), the tree takes $1.08\times$ the time of bCSR, filling $2.55\times$ its memory; the host-link bound, $0.48\times$ our PageRank, excludes no faster engine.
\end{enumerate}

\section{Related Work}\label{sec:related}
Sums and products of \ktree{}s, and the RPQs built on them, run on the CPU with one dependent $\mathrm{rank}$ per visited node~\cite{rpq-k2tree-vldbj,time-space-rpq}; depth-first~\cite{carmona-depth-first-2025} and depth-first unary degree sequence~\cite{navarro-cache-friendly-2026} layouts, borrowed from succinct ordinal trees~\cite{Munro_Raman_2001,jansson2007ultra,navarro2014fully}, improve their locality. Pointer-based quadtrees~\cite{samet2006foundations} with CSR or dense leaves store valued sparse matrices, but their control overhead makes SpMV slower than CSR~\cite{spmv-quadtree}; the \ktree{} is their succinct, pointerless form, and two-dimensional block trees further share repeated submatrices~\cite{2DBT}. Multi-terminal BDDs~\cite{bryant1986graph_based_boolean,fujita1997multiterminal,bahar1997decision_diagram} split a matrix into quadrants as the \ktree{} does, share identical ones, and multiply on them. RPQs, central to graph query languages~\cite{modern_query_lang_graph_db,cypher,PGQL,GQL}, run over compact indexes~\cite{rpq-compact-vldbj}, as sparse Boolean linear algebra~\cite{LA-RPQ}, also over \ktree{}s~\cite{rpq-k2tree-vldbj} and qdags~\cite{arroyuelo2026boosting}, and on GPUs over uncompressed graphs~\cite{cuRPQ}. Direction-optimizing search~\cite{beamer2013direction,push_pull_graphblas} pushes from sparse frontiers, whereas our products traverse the whole tree, which suits dense vectors and batches. Out-of-memory GPU graph systems cut the CSR bytes that cross the host link~\cite{subway,ascetic,HyTGraph,grus,HALO,emogi}, and GCGT~\cite{gpu_traversal_compressed_graphs} decodes Boldi--Vigna-style adjacency lists on the GPU beyond device memory; the tree is read in place, serves both directions and leaves the link idle. The grammar engine~\cite{c5} has the level-synchronous, double-buffered structure we adopt, over a RePair grammar~\cite{repair,repair_journal}. CSR-DU~\cite{CSR-DU} delta-encodes the column indices of a CSR on multicore CPUs, CSX~\cite{CSX} also its runs and blocks, and BRO~\cite{BRO} packs them on GPUs; all keep a value per non-zero. Numerical work on memory-bound kernels shortens the indices and values of sparse matrices~\cite{aliaga2021balanced,g-ANS} or stores values, such as a Krylov basis, in fewer bits than they are computed in~\cite{ginkgo_memory_accessor,frsz2}; we take the Boolean, lossless extreme, the memory corner of the read--update--memory trade-off~\cite{RUM-conjecture}. GPU engines for iterative relational algebra and Datalog compute closures over uncompressed relations as fixed points of joins~\cite{shovon2023iterative,gpu_datalog}.
Blocked formats exploit the two-dimensional structure that the \ktree{} recurses on. Compressed sparse blocks (CSB)~\cite{CSB} store the non-zeros of each block in Z-Morton order, so that $Ax$ and $A^{\mathsf{T}}x$ run in parallel from one structure, as our right and left products do; like HiCOO for sparse tensors~\cite{hicoo}, CSB keeps an entry and a value per non-zero on multicore CPUs, where the \ktree{} keeps one bit per non-empty block at each level. On GPUs, TileSpMV~\cite{tile_spmv} selects a format per $16\times16$ tile and DASP~\cite{dasp} runs SpMV on the matrix-multiply-accumulate units; both store values. Bit-GraphBLAS~\cite{bit_graphblas} is the closest: its B2SR format stores the non-empty tiles of a Boolean matrix as dense bit matrices under a block CSR, so a tile costs its full size even for one non-zero, and the transpose needs a second, CSC index; the \ktree{} recurses down to $2\times2$ leaves and serves rows and columns alike. Morton codes and scans also build spatial trees on GPUs~\cite{karras}.

\section{Conclusions}\label{sec:concl}
A \ktree{} can be multiplied efficiently on a GPU once its traversal is organized by levels: the traversal does not depend on the vector, so the child offsets that a descent obtains by dependent $\mathrm{rank}$ lookups become static data, scanned at each product or read from a sampled directory, and tiles bound the memory of a product by the depth of the tree. The resulting engine computes right and left products, single or batched, over three semirings, from a structure $2.3$--$27\times$ smaller than the cuSPARSE CSR, within a small factor of cuSPARSE in time. Its fewer bytes save time where memory bandwidth is scarce, as on the \GB{}, and buy capacity on any GPU: on an H100, it multiplies a 92-billion-edge web crawl beyond device memory, and runs PageRank on it, faster than any streamed CSR. Two directions remain: streaming the tiles of a tree that itself exceeds device memory, and products of two trees~\cite{rpq-k2tree-vldbj,arroyuelo2026boosting}, whose frontier of node pairs would compute Boolean $A\cdot B$ and, by repeated squaring, a compressed transitive closure.

\begin{acks}
All the H100 computations presented in this paper were performed using the GRICAD infrastructure (\href{https://gricad.univ-grenoble-alpes.fr}{https://gricad.univ-grenoble-alpes.fr}), which is supported by Grenoble research communities. Francesco Tosoni was funded by the Alfred P. Sloan Foundation with the grant \#\href{https://sloan.org/grant-detail/g-2025-25193}{G-2025-25193} (\url{sloan.org}). Francesco Tosoni has received travel support from SWH (Inria Paris) and the Wikimedia Foundation; the SWH archive, Wikidata and Wikipedia are among the data sources of this paper.
\end{acks}

\bibliographystyle{ACM-Reference-Format}
\bibliography{references}

\end{document}